# Topology optimization of simultaneous photonic and phononic bandgaps and highly effective phoxonic cavity

Hao-Wen Dong,[1] Yue-Sheng Wang,[1,*] Tian-Xue Ma,[1] and Xiao-Xing Su[2]

[1]*Institute of Engineering Mechanics, Beijing Jiaotong University, Beijing 100044, China*
[2]*School of Electronic and Information Engineering, Beijing Jiaotong University, Beijing 100044, China*
*\*Corresponding author: yswang@bjtu.edu.cn*

By using the non-dominated sorting-based genetic algorithm II, we study the topology optimization of the two-dimensional phoxonic crystals (PxCs) with simultaneously maximal and complete photonic and phononic bandgaps. Our results show that the optimized structures are composed of the solid lumps with narrow connections, and their Pareto-optimal solution set can keep a balance between photonic and phononic bandgap widths. Moreover, we investigate the localized states of PxCs based on the optimized structure and obtain structures with more effectively multimodal photon and phonon localization. The presented structures with highly focused energy are good choices for the PxC sensors. For practical application, we design a simple structure with smooth edges based on the optimized structure. It is shown that the designed simple structure has the similar properties with the optimized structure, i.e. simultaneous wide phononic and photonic bandgaps and a highly effective phononic/photonic cavity, see Figures 8(b) and 8(c).

## 1. Introduction

Introduction of periodicity into structures leads to the so-called phononic crystals (PnCs) [1] and photonic crystals (PtCs) [2, 3]. Due to the possible full bandgap characteristic, PnCs can be designed by selecting physical and geometrical parameters for different and surprising potential applications, e.g., damping [4, 5], isolation [6], rectification [7], confinement [8] and guiding [9] of acoustic or elastic waves. In addition, based on the design of equal-frequency contours of PnCs, the negative refraction [9], focusing [11], beam splitting [12], self-collimation [13, 14], and acoustic diodes [15, 16] can be obtained. Similarly, in the context of PtCs, many studies focus on the designing for controlling the propagation of electromagnetic waves in specific wavelength ranges. PtCs also have many promising applications, e.g., optical sensors [17], telecommunications [18], fibers [19] and wavelength multiplexers/de-multiplexers [20]. Obviously, how to better control, localize, and guide the sound (acoustic waves) and light (electromagnetic waves) simultaneously is an appealing research area.

In recent years, many scientists have investigated the simultaneous control of the acoustic and optical properties in the same system based on phoxonic crystals (PxCs) [21-27]. PxCs, also known as phononic-photonic or optomechanical structures, have dual bandgaps for both photons and phonons simultaneously leading to quite promising applications, i.e., enhancing the acousto-optic or optomechanical interactions [28-32], effectively manipulate photons with phonons [21, 22, 24] and co-localized photonic and phononic resonances [33, 34]. Many researchers have investigated the opening of simultaneous photonic and phononic bandgaps [21-24, 26, 27] and acousto-optic interaction [28-32] by using geometrical parameters. For one-dimensional structures, Trigo et al. [35] observed the cavity-confined high frequency hyper-sound mode within an optical microcavity based in GaAs and AlAs materials. Lachrmoise et al. [36] described a device that has a resonant cavity for acoustic phonons embedded inside an optical cavity and designed the material parameters for the optimization of the acoustic phonon cavities. Psarobas et al. [37] reported the occurrence of strong nonlinear acoustic-optic interactions in phoxonic cavity. Papanikolaou et al. [38] have demonstrated that, in a phoxonic nanostructure, localized phonons can trigger nonlinear acousto-optic interactions leading to strong modulation of light with sound. As for two-dimensional (2D) cases, several structures are investigated to obtain simultaneous photonic and phononic bandgaps [21, 24, 39, 40], cavities [22, 42] and waveguides [42] in PxCs. Maldovan and Thomas [18] have reported that triangle-latticed structures with air rods in an elastic matrix can obtain large, complete and simultaneous photonic and phononic bandgaps. They also demonstrated the simultaneous localization of photons and phonons in the same spatial region by introducing lattice defects in PxCs [22]. Sadat-Saleh et al. [24] observed that decreasing the symmetry of the lattice by adding atoms of different size inside the unitcellunitcell leads to larger phoxonic bandgaps. Bria et al. [39] also investigated the opening of phoxonic bandgaps in a simple square periodic array of holes drilled in sapphire and silicon substrates. Ma et al. [40] proposed the PxCs with veins in square, triangle and honeycomb lattices and got large phoxonic complete bandgaps. In addition, due to the low loss as the energy can be well confined within the thickness of the slab, studies of 2D phoxonic slabs with different topologies [26, 27, 30, 41, 42] have been performed for simultaneous phononic and photonic bandgaps and optomechanical cavities and waveguides [42]. Rolland et al. [28] examined the acoustic-optic coupling in PxCs based on both photoelastic and optomechanical mechanisms. Lucklum et al. [43] showed that the PxC slab with a defect can provide very high sensitivity for a PxC sensor platform. For three-dimensional (3D) cases, Akimov et al. [23] studied the experimental evidence of existence of simultaneous bandgap in 3D structure. The opening of phoxonic bandgap in a metallodielectric cubic lattice [25] and a simple cubic lattice [44] was reported, respectively.

Obviously, PxCs have promising future in the control and management of the elastic, acoustic and electromagnetic waves. However, we have to take full advantages of PxCs when applying them to the real world. So, how to design PxCs with best performance for different applications becomes a very crucial topic. The above studies are all based on the particular geometrical topology and materials. This is obviously inadequate for getting the PxC devices with good enough properties. Fortunately, in the quest for an optimal structure, the topology optimization provides a powerful means and shows its significance in the area of designing PtCs [45-52] and PnCs [53-59]. As far as the present authors know there is no investigation on the topology optimization of PxCs or PxC devices yet.

In this paper, by using the non-dominated sorting-based genetic algorithm II (NSGA-II) [60], we study the topology optimization of 2D PxCs with simultaneously large and complete photonic and phononic bandgap widths. Based on the optimized structure, we study the optimization of PxC cavities using NSGA-II. We show that a structure with more effectively multimodal photon and phonon localization can be obtained. This structure exhibits highly focused energy and is a good choice for the PxC sensors [43]. The band structures and defect modes are computed by using the finite element method (FEM) for a 2D square lattice porous crystal. We first describe the multi-objective optimization problems of PxCs. Then, the optimized structures with extremely large bandgap and the optimized PxC cavities with highly focused energy are presented, respectively. Finally, some general conclusions followed by the discussions are presented.

## 2. Topology optimization of PxCs

To design a 2D PxC with simultaneously maximal complete photonic and phononic bandgaps, it is shown that the objectives are conflicting, i.e., the improvement of one objective will lead to the deterioration of the others [40, 44]. So, to obtain the maximal photonic and phononic bandgaps simultaneously, we use the NSGA-II [60] which is the most popular and efficient optimization algorithm to perform the multi-objective optimization of PxCs. In topology optimization, a structure is optimized by removing and adding elements in the unitcell (i.e., the design domain) to obtain an optimized porous structure. Thus, the optimization problem herein with the constraint is taken as

Maximize:
$$C_{pt}(\Sigma) = \max_{\forall n}\left\{2*\frac{\min_{\mathbf{k}}:\omega_{n+1}^{pt}(\Sigma,\mathbf{k})-\max_{\mathbf{k}}:\omega_{n}^{pt}(\Sigma,\mathbf{k})}{\min_{\mathbf{k}}:\omega_{n+1}^{pt}(\Sigma,\mathbf{k})+\max_{\mathbf{k}}:\omega_{n}^{pt}(\Sigma,\mathbf{k})}\right\}, (1)$$

Maximize:
$$C_{pn}(\Sigma) = \max_{\forall n}\left\{2*\frac{\min_{\mathbf{k}}:\omega_{n+1}^{pn}(\Sigma,\mathbf{k})-\max_{\mathbf{k}}:\omega_{n}^{pn}(\Sigma,\mathbf{k})}{\min_{\mathbf{k}}:\omega_{n+1}^{pn}(\Sigma,\mathbf{k})+\max_{\mathbf{k}}:\omega_{n}^{pn}(\Sigma,\mathbf{k})}\right\}, (2)$$

Subject to: $\rho_i = 0 \text{ or } 1$, (3)

$$\min_{\Sigma}(p) \geq p^*, (4)$$

$$C_{pt} \geq 0.1, (5)$$

where $\Sigma$ denotes the topological distribution within the unitcell composed of the vacuum and silicon; $C_{pt}$ is the relative complete bandgap width (BGW) for combined transverse electric (TE) and transverse magnetic (TM) modes; $C_{pn}$ is the relative complete BGW for combined out-of-plane and in-plane wave modes; and $n$ denotes the serial number of the energy bands (in this paper, we take $n$=1, 2, …9). $\rho_i$ is the mass density of the element and declares the absence (0) or presence (1) of an element. In particular, to control the geometrical width of every place of the topology so that the optimized structure are easy for fabrication, we introduce a geometrical constraint in Eq. (4), that is, the geometrical width $p$ of every connection must be bigger than $p^*$ which is selected as $a$/30 in this paper.

What we should note here is that the constraint in Eq. (5) is not necessary in the optimization problem. However, we have to use this constraint herein based on the following two reasons. According to the previous studies [21, 24, 39, 40], we find that it is harder to open a large enough complete photonic bandgap than to open a large phononic bandgap. Besides, with the limited population size we used, our numerical tests suggest that only a few solutions with large $C_{pt}$ and $C_{pn}$ can be obtained due to the distribution of the limited solutions in the search space. Therefore we set a constraint of $C_{pt} \geq 0.1$ to ensure getting the effective solutions with the limited computational cost.

To simplify calculation, we assume that the primitive unitcell has a square-symmetry. Thus only one-eighth of the whole unitcell really needs to be determined. And the Bloch wave vector can only be selected at the boundaries of the irreducible Brillouin zone. However, we are not sure that the system with this symmetry can yield a better solution [57]. Therefore, the topology optimization of PxCs without assumption of any symmetry is an interesting topic and will be considered in the future.

In this paper, the optimization for the 2D square-latticed porous PxCs with anisotropic silicon is performed. The material parameters are: the refractive index $n$=3.6, mass density $\rho$=2331 kg/m$^3$, and elastic constants $C_{11}$=16.57$\times 10^{10}$ Pa, $C_{12}$=6.39$\times 10^{10}$ Pa and $C_{44}$=7.962$\times 10^{10}$ Pa [40]. The algorithm parameters of the NSGA-II [60] are the population size $N_p$=30, the crossover probability $P_c$=0.9 and the mutate probability $P_m$=0.02. We firstly obtain the optimized solutions based on coarse 30×30 pixels [57-59] after the 1000 evolutionary generations. Then, in order to obtain the optimized structures with smooth edges, all obtained optimized structures are mapped to finer 60×60 grids, and then are used as the initial population of the new run of the NSGA-II [60]. The numerical tests show that, after the fixed maximal generation of 2000, the final optimization results are produced.

Figure 1 illustrates the multi-objective optimization results of the problem in Eqs. (1)-(5). According to Fig. 1, all optimized solutions are the feasible solutions, i.e., the values of their objective function $C_{pt}$ are larger than or equal to 0.1. It is seen from Fig. 1 that $C_{pn}$ and $C_{pt}$ are opposite for optimization of PxCs, i.e., the improvement of one objective will lead to the deterioration of the other. Every Pareto-optimal solution affords the non-dominated combination of $C_{pn}$ and $C_{pt}$. The decision makers can select one non-dominated combination on the basis of requirement. If $C_{pt}$ is more cared, the non-dominated solutions in the upper side of the curve are good choices. Conversely, if $C_{pn}$ is more cared, the solutions in the lower side are more excellent. As shown in Fig. 1, all the structures of the Pareto-optimal solutions have solid lumps with narrow connections, just like the reported structures in Refs. [40], [44], [59] and [61]. In particular, the phononic and photonic band structures of the design E in Fig. 1 are presented in Fig. 2. Obviously, both phononic and photonic relative BGWs of 0.723 and 0.141 are large enough. However, it is difficult to simultaneously open so ultrawide phononic and photonic bandgaps by selecting physical and geometrical parameters [21, 22, 24, 39, 40]. Therefore, the multi-objective topology optimization is very effective for designing PxC band structures. And, it is expected to be applied in design of acoustic-optical or optical-acousitcal devices [22]. In the following section, we will select the design E as the based structure to perform the topology optimization of the defect of the supercell for designing the PxC cavity with highly focused energy.

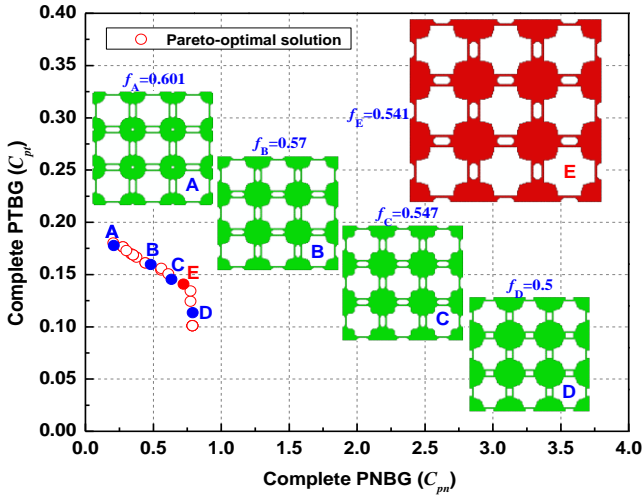

Fig. 1. Optimized solutions for the maximal relative complete BGW of the acoustic and optical modes simultaneously. The representative structures (A, B, C, D and E) of the Pareto-optimal solutions are shown, respectively. The corresponding porosities ($f$) of these five structures are also shown.

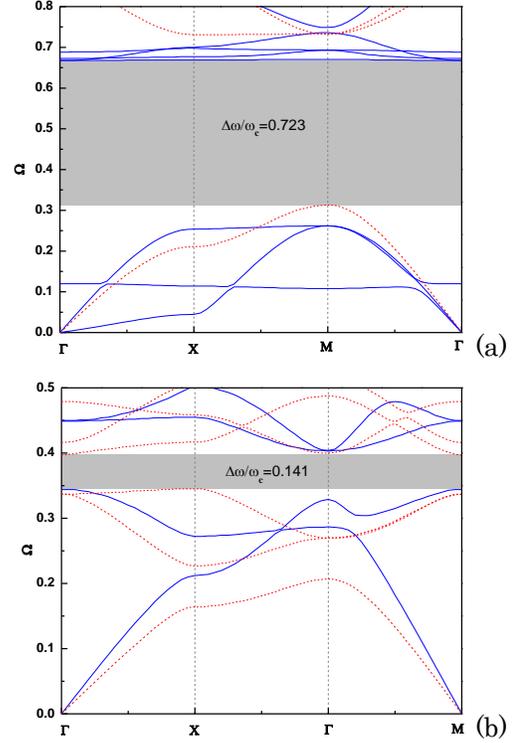

Fig. 2. Pnononic (a) and photonic (b) bandgap structures of the design E in Fig. 1. The solid and dotted lines in (a) represent the in-plane and the out-of-plane wave modes, respectively. The solid and dotted lines in (b) represent the TM and the TE modes, respectively.

### 3. Design of the PxC cavity with highly focused energy

Just like the simultaneous localization of photons and phonons by introducing lattice defects in air-silicon PxCs presented by Maldovan et al. in 2006 [22], the design E with a point defect also can locate the simultaneous electromagnetic and elastic states around the defect. Nevertheless, the spatial distributions of the electromagnetic and elastic fields are generally very fragmented. Can we design a special point defect to get the highly focused spatial distributions of the electromagnetic and elastic fields simultaneously? For designing the PxCs with this property, we also can use the topology optimization to realize it based on the original supercell composed of the design E. We introduce two specific parameters, i.e., $D_{pt}$ and $D_{pn}$, to represent the energy concentration extent for the defect state. The energy concentration extent is decided by computing the distance between the center of the gravity of the spatial distributions of the one-eighth of the design domain and the central point of the design domain. Closer these two objective points are, more centralized energy will be achieved. So, the topology optimization problem is stated as

Minimize:

$$D_{pt}(\Sigma) = \min_{\forall n}\left\{\sqrt{\left(\frac{\int_0^{a/2}\int_0^y xf_n^{pt}(x,y)dxdy}{\int_0^{a/2}\int_0^y f_n^{pt}(x,y)dxdy}-x_0\right)^2+\left(\frac{\int_0^{a/2}\int_0^y yf_n^{pt}(x,y)dxdy}{\int_0^{a/2}\int_0^y f_n^{pt}(x,y)dxdy}-y_0\right)^2}\right\}$$
, (6)

Minimize:

$$D_{pn}(\Sigma) = \min_{\forall n}\left\{\sqrt{\left(\frac{\int_0^{a/2}\int_0^y xf_n^{pn}(x,y)dxdy}{\int_0^{a/2}\int_0^y f_n^{pn}(x,y)dxdy}-x_0\right)^2+\left(\frac{\int_0^{a/2}\int_0^y yf_n^{pn}(x,y)dxdy}{\int_0^{a/2}\int_0^y f_n^{pn}(x,y)dxdy}-y_0\right)^2}\right\}$$
, (7)

Subject to: $\rho_i = 0$ or $1$, (8)

$\min_{\Sigma}(p) \geq p^*$, (9)

$D_{pt} \leq 0.0055$, (10)

where $\Sigma$ denotes the topological distribution within the design domain of the point defect in Fig. 3; $D_{pt}$ and $D_{pn}$ are the minimal distances of the photonic and phononic defect states, respectively; $n$ denotes the serial number of the existent defect states; $a$ is the length of the design domain; $x$ and $y$ are the $x$-coordinate and $y$-coordinate values of the node, respectively; $f_n^{pt}(x,y)$ and $f_n^{pn}(x,y)$ are the electromagnetic and displacement fields values of the $n$th defect state, respectively; $(x_0, y_0)$ is the center coordinate of the design domain in Fig. 3. According to Eqs. (6)-(10), the optimized defect with a simultaneously highly-centralized localization of photons and phonons is likely to be found.

Just like the constraint in Eq. (5), the use of the constraint in Eq. (10) is also based on the following two facts. On one hand, our numerical tests suggest that it is harder to get the solutions with small $D_{pt}$ than those with small $D_{pn}$. On the other hand, with the limited population size we used, only a few solutions with small $D_{pt}$ and $D_{pn}$ can be obtained due to the distribution of the limited solutions in the search space. Thus we set a constraint of $D_{pt} \leq 0.0055$ to ensure getting as many solutions with small enough $D_{pt}$ and $D_{pn}$ as possible with the limited computational cost.

To simplify the calculation, we assume that the defect has a square-symmetry. Thus only one-eighth of the whole design domain really needs to be determined. We also use the NSGA-II [60] to solve the problem of Eqs. (6)-(10). In fact, if considering both the TE and TM modes for the objective function $D_{pt}$ in Eq. (6) and considering both the in-plane and out-of-plane wave modes for the objective function $D_{pn}$ in Eq. (7), it is very difficult to perform the multi-objective optimization, not to mention how hard it will be to obtain the Pareto-optimal solution set with highly focused energy for simultaneous optical and acoustic cavity modes. So, to reach this goal and reduce the search space, we first solve the multi-objective optimization problem just considering the in-plane acoustic and the TE optical cavity modes.

The 5×5 original supercell with the design domain and the Pareto-optimal solutions are presented in Fig. 3. It is shown that $D_{pn}$ and $D_{pt}$ are mutually exclusive in the optimization problem of Eqs. (6)-(10). We choose a solution $O_1$ at the front of the Pareto-optimal solution set and show the corresponding optimized supercell and the optimized defect. Like the design E in Fig. 1, the optimized defect also has solid lumps with narrow connections [40, 44, 59, 61]. Figures 4(a, b) and 4(c, d) displays the field distributions of the localized states for phonons and photons, respectively. The magnitudes of the corresponding field vectors are expressed by the color scheme. The normalized cavity frequencies of the four modes in Fig. 4 are 0.32, 0.314, 0.376 and 0.352, respectively. It is noted that the out-of-plane acoustic and the TM optical cavity modes in Figs. 4(b) and 4(d) correspond to the defect which is optimized based on the in-plane acoustic and the TE optical cavity modes. Although, in the topology optimization, the out-of-plane acoustic and the TM optical modes are not under consideration. The results in Figs.4 (b) and 4(d) show that the cavity modes can still have small $D_{pn}$ and $D_{pt}$. From Fig. 4 we note that, if both the elastic and electromagnetic waves with the appropriate frequencies incident onto the structure $O_1$, they all will be localized in the lump in the center of the optimized defect. Moreover, for all four modes, the large magnitudes just occur in the same area of the optimized supercell. This means that ultra-strong photon-phonon interactions can be motivated through the optimized structure. If we use this optimized structure as the PxC sensors [43], a high sensitivity will be expected because of the high energy concentration.

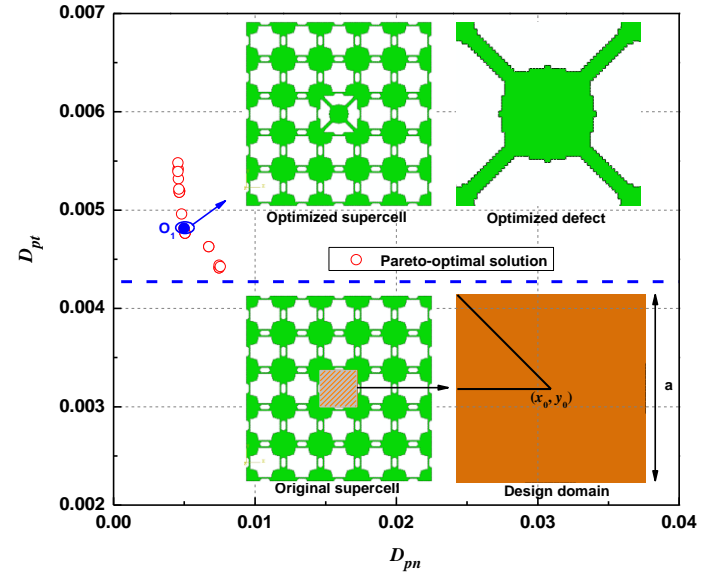

Fig. 3. Multi-objective optimization results with consideration of the in-plane acoustic and the TE optical cavity modes. The 5×5 original supercell with the design domain, the Pareto-optimal solutions, the optimized supercell and its optimized defect are shown for the representative design $O_1$ of the Pareto-optimal solution set. Only one-eighth of the whole design domain, i.e., the triangular area, really needs to be determined.

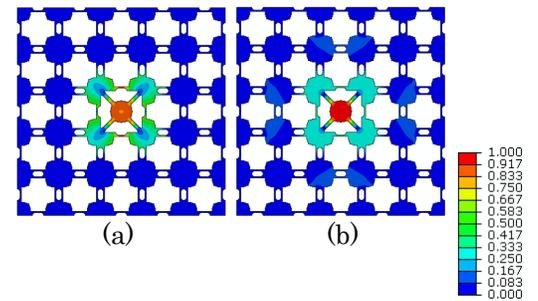

(a) (b)

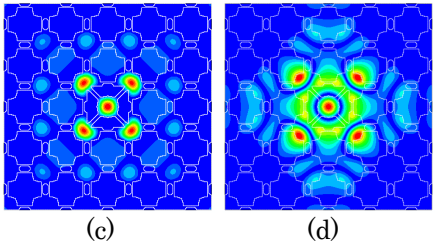

(c)   (d)

Fig. 4. The acoustic and optical cavity modes of the representative design $O_1$: (a) Spatial distribution of the magnitudes of the displacement fields for the in-plane acoustic cavity mode with the minimal $D_{pn}$; (b) Spatial distribution of the magnitudes of the displacement field for the out-of-plane cavity mode; (c) Electric-field distribution of the magnitudes for the TE mode with the minimal $D_{pt}$; and (d) Magnetic-field distribution of the magnitudes for the TM mode. Note that, the out-of-plane acoustic and the TM optical cavity modes correspond to the defect which optimized is based on the in-plane acoustic and the TE optical cavity modes.

In order to validate the optimized results in Figs. (3) and (4), we present in Fig. 5 the eigenfrequencies of the cavity modes in the band diagrams together with the corresponding mode profiles, respectively, for the acoustic and optical fields. For clarity, all defect state frequencies are represented by the solid lines in the bandgap regions whose upper and lower band edges are showed by the dash lines. Several modes are observed within the bandgaps. Especially, the lines *a*, *b*, *c* and *d* correspond to the cavity frequencies of the modes in Figs. 4(a), 4(b), 4(c) and 4(d). For comparison, we show in Fig. 5(e) the defect modes (i.e. na, nb, nc and nd) whose eigenfrequencies are closest to those of the modes *a*, *b*, *c* and *d*. It is seen that the energy concentration extent of the representative cavity modes in Fig. 5(e) are smaller than that of the optimized cavity modes in Fig. 4. That is, the presented optimization method can find out the optimized results accurately.

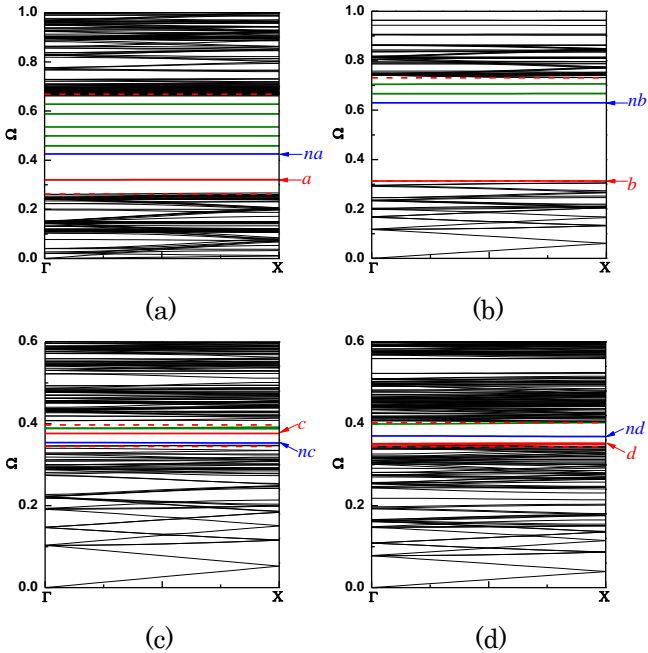

(a)   (b)
(c)   (d)

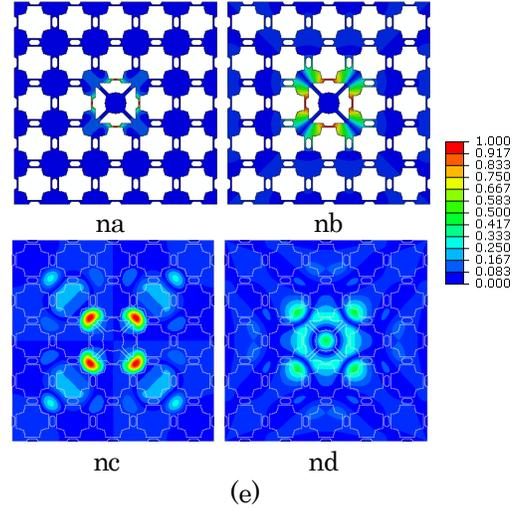

na   nb
nc   nd
(e)

Fig. 5. Defect modes within bandgaps and the field distributions for the defect modes near the optimized cavity modes: (a) in-plane acoustic mode; (b) out-of-plane acoustic mode; (c) TE polarized optical mode; (d) TM polarized optical mode; and (e) the field distributions for the corresponding cavity mode.

In addition, we also investigate the multi-objective optimization problem with consideration of the out-of-plane acoustic and the TE optical cavity modes. The multi-objective optimization results including the Pareto-optimal solution set, the optimized supercell and its optimized defect are presented in Fig. 6. It is shown that the Pareto-optimal solution set provides many effective choices and keeps a balance between the two objective functions. Moreover, the optimized defect of the representative design $O_2$ has similar geometric feature as the design $O_1$ in Fig. 3. That is, it is also composed of the solid lumps with narrow connections despite of the difference on the edges of lumps and connections. In order to validate the optimized structure, the acoustic and optical cavity modes with the minimal $D_{pt}$ and $D_{pn}$ are plotted in Fig. 7. The normalized cavity frequencies of the four modes in Fig. 7 are 0.315, 0.314, 0.376 and 0.35, respectively. Obviously, similar to the design $O_1$, the design $O_2$ also has highly focused energy for the acoustic and optical cavity modes simultaneously. Based on the similarity between Figs. 4 and 7, we can say that it is better to design the PxC cavity with consideration of the in-plane (or the out-of-plane) acoustic and the TE optical cavity mode. Besides, according to Figs. 3 and 6, the PxCs composed of the solid lumps with narrow connections are well suitable for optomechanical waveguides and microcavities. On one hand, the PxCs with this geometric property can open ultrawide PxC bandgaps. One the other hand, the defect with this geometric property can produce ultrastrong photon-phonon interactions in the same area.

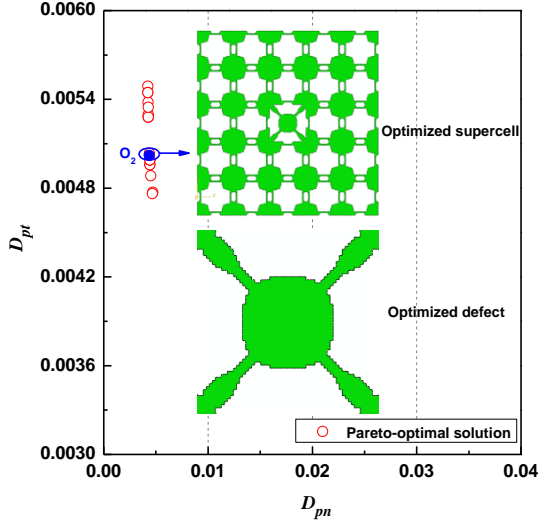

Fig. 6. Multi-objective optimization results with consideration of the out-of-plane acoustic and the TE optical cavity modes. The optimized supercell and its optimized defect of the representative design $O_2$ of the Pareto-optimal solution set are shown.

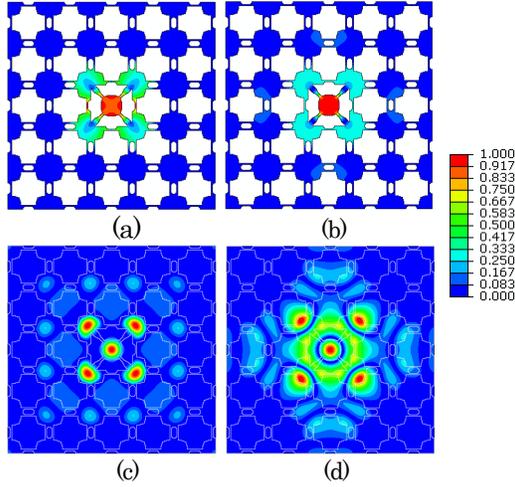

Fig. 7. The acoustic and optical cavity modes of the representative design $O_2$: (a) Spatial distribution of the magnitudes of the displacement fields for the in-plane acoustic cavity mode; (b) Spatial distribution of the magnitudes of the displacement fields for the out-of-plane cavity mode with the minimal $D_{pn}$; (c) Electric-field distribution of the magnitudes for the TE mode with the minimal $D_{pt}$; and (d) Magnetic-field distribution of the magnitudes for the TM mode. Note that, the in-plane acoustic and the TM optical cavity modes correspond to the defect which optimized is based on the out-of-plane acoustic and the TE optical cavity modes.

## 4. Conclusions

We successively perform the topology optimization of the 2D PxCs with simultaneously maximal and complete photonic and phononic bandgap widths and the PxC cavity, respectively. The multi-objective topology optimization shows its effectiveness for designing PxCs band structures. And, it is expected to be extended to design the acoustic-optical and optical-acousitcal devices. For reducing the computational difficulty and increasing the chances for finding the optimized solutions, it is better to design the PxC cavity with consideration of the in-plane (or the out-of-plane) acoustic and the TE optical cavity mode. Moreover, due to the ultrawide PxC bandgaps and the ultrastrong photon-phonon interactions, the presented optimized PxCs composed of the solid lumps with narrow connections are well suitable for optomechanical waveguides and microcavities.

In spite of the complexity of the optimized structures in this paper, the optimized solutions can reveal the beneficially geometrical properties for pursuing the acoustic and optical devices and thus are important inspiration for designing the corresponding devices. For example, we can design a relatively simple structure with smooth edges based on the optimized structure $O_1$. Fig. 8(a) shows the realistically designed size of the PxC structure where $a=0.5a$, $b=0.1a$, $w_1=0.07a$, $\theta_1=13.5°$, $c=0.033a$, $d=0.067a$, $e=0.8a$, $w_2=0.033a$, and $\theta_2=45°$. The lattice constant is $a=200$ nm. The phononic and photonic relative BGWs are 0.756 and 0.128, respectively. The cavity frequencies of the four modes in Fig. 8(c) are 9.171 GHz, 8.86 GHz, 525.126 nm and 564.978 nm, respectively. Obviously, the simple structure has the similar properties with the structure $O_1$.

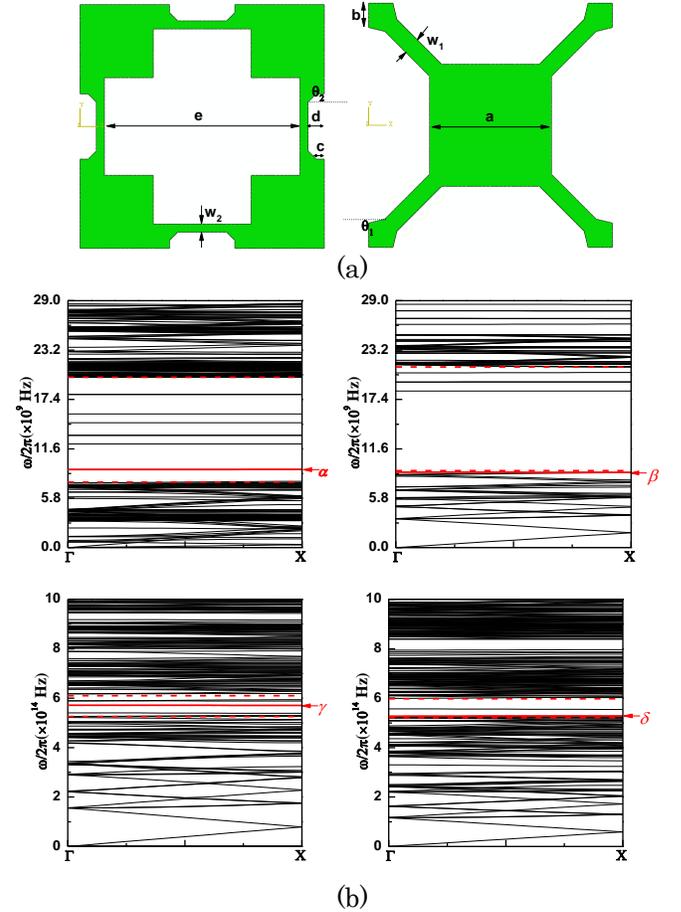

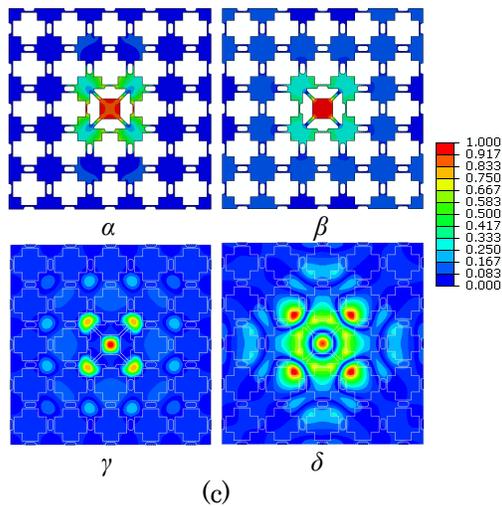

(c)

Fig. 8. (a) The simple unitcell and defect with smooth edges suggested by the design $O_1$. (b) The band structures of the 5×5 PxC supercell composed of the unitcell and defect in (a). (c) The field distributions for the corresponding cavity modes: the in-plane acoustic mode $\alpha$; the out-of-plane acoustic mode $\beta$; TE polarized optical mode $\gamma$; and TM polarized optical mode $\delta$.

## Acknowledgements

This research is supported by the National Natural Science Foundation of China (Grant Nos. 11372031), and partially by the National High Technology Research and Development Program of China (863 Program) (Grant No. 2013AA030901). The second author is also grateful to the support of the Fundamental Research Funds for the Central Universities under Grant No. 2013 JBM009.

J. Appl. Phys. **110**, 113520 (2011).